\documentclass[prl,a4paper,twocolumn,showpacs,superscriptaddress,floatfix]{revtex4}
\usepackage{times}
\usepackage{graphicx}
\usepackage{amsmath}
\usepackage{amssymb}
\usepackage{bm}
\newcommand{\be}{\begin{equation}}
\newcommand{\ee}{\end{equation}}

\begin{document}

\title{Correlation functions and correlation widths in Quantum-Chaotic Scattering for mesoscopic systems and nuclei}

\author{J. G. G. S. Ramos}
\affiliation{Departamento de F\'isica, Universidade Federal da Para\'{\i}ba, 58051-970, Jo\~ao Pessoa, Para\'iba, Brazil}

\author{A. L. R. Barbosa}
\affiliation{
 Departamento de F\'isica, Universidade Federal Rural de Pernambuco, Dois Irm\~aos, 52171-900 Recife, Pernambuco, Brazil}
\author{B. V. Carlson} 
\affiliation{Departamento de F\'{i}sica, Instituto Tecnol\'{o}gico de Aeron\'{a}utica, CTA, S\~{a}o Jos\'{e} dos Campos, S.P., Brazil.}
\author{T. Frederico}
\affiliation{Departamento de F\'{i}sica, Instituto Tecnol\'{o}gico de Aeron\'{a}utica, CTA, S\~{a}o Jos\'{e} dos Campos, S.P., Brazil.}
\author{M. S. Hussein}
\affiliation{Instituto de Estudos Avan\c cados and Instituto de F\'{\i}sica, Universidade de S\~{a}o Paulo,
  C.P.\ 66318, 05314-970 S\~{a}o Paulo, SP, Brazil.\\
  Departamento de F\'{i}sica, Instituto Tecnol\'{o}gico de Aeron\'{a}utica, CTA, S\~{a}o Jos\'{e} dos Campos, S.P., Brazil.}

\date{\today}

\begin{abstract}

We derive analytical expressions for the correlation functions of the electronic conductance fluctuations of an open quantum dot under several conditions. Both the variation of energy and that of an external parameter such as an applied perpendicular or parallel magnetic fields are considered in the general case of partial openness. These expressions are then used to obtain the ensemble averaged density of maxima, a measure recently suggested to contain invaluable information concerning the chaoticity of the system. The correlation width is then calculated for the case of energy variation and a significant deviation from the Weisskopf estimate is found in the case of two terminals. The results are extended to more than two terminals. All our results are analytical.The use of these results in other fields, such as nuclei, where the system can only be studied through a variation of the energy, is then discussed. 
\end{abstract}


\maketitle

\section{Introduction}
Chaotic quantum systems, such as open Quantum Dots \cite{Alhassid}, graphene flakes \cite{flakes1,flakes2,flakes3}, and nuclei exhibit common universal features characterized by fluctuating observables, such as the electronic conductance in the former two, and the compound nucleus cross section in the latter \cite{weiden1,weiden2,weiden3}. Useful information about the system in this regime of the dynamics are obtained through the average of the observables and through the correlation functions defined as the average of products of two fluctuating observables at two values of the independent variable, such as the energy or an external field. The correlation function is of paramount importance as it measures the degree of coherence present in the otherwise fully chaotic system. This measure resides in the correlation width which specifies the shape of the correlation function. In the case of parametric energy variation the shape is Lorentzian, as shown more than half a century ago by Ericson \cite{Ericson}, while in the case of a varying external parameter such as a magnetic field, the shape is a square Lorentzian, \cite{Efetov,Vallejos, caio}. In \cite{caio}, the effect of temperature on the conductance correlation function was investigated. Deviations from these shapes have also been investigated \cite{ref2,caio}. For their importance in the study of chaotic mesoscopic systems, nanosystems, resonators and in nuclei, it is clear that a detailed investigation of the correlation functions and their potential deviations from the above mentioned two extreme shapes and the resulting correlation widths is called for. In this paper we calculate analytically the correlation functions for the general case of partially open system. Three types of external magnetic fields are considered. The energy correlation function is discussed in details and the corresponding correlation width is  extracted and found to be significantly different from the Weisskopf estimate. This last result is quite important as it affects the value of the dwell time in these systems.

\section{Quantum Chaotic Scattering} 
Quantum Chaotic Scattering (QCS) is a widely occurring phenomenon in physics. It operates in a variety of systems, such as, electronics \cite{Alhassid}, spintronics \cite{spintronica}, biomolecules \cite{bio}, disordered mesoscopic nanostructutes \cite{Beenakker}, and the compound nucleus \cite{Bohr}. The emergence of the phenomenon is, just as in other cases of quantum chaos, directly related to the intrinsic chaotic dynamics associated with quasi-bound states of a quantum system. Random Matrix Theory (RMT) supplies the formal S-matrix  of QCS, as it describes the intrinsic Hamiltonian which governs the dynamics and the scattering. The empirical manifestation of QCS is evidenced by universal fluctuations observed in electronic conductance in open Quantum Dots, and in Graphene flakes, transmittance in microwave and acoustic resonators, and in compound nucleus cross sections. \\

The S-matrix describing QCS is given by

\begin{equation}
\label{eq:SHeidelberg}
S (\varepsilon, X) = \mbox{$\openone$} - 2\pi i W^\dagger (\varepsilon - H(X) + 
i \pi W W^\dagger)^{-1} W \;,
\end{equation}

where $H(X)$ is a random Hamiltonian matrix of dimension $M \times M$ that describes the resonant states in the chaotic system, which is subject to the influence of an external parameter, X. The number of resonances is very large ($M \rightarrow \infty$). The matrix $W$ of dimension $M \times (N)$ contains the channel-resonance coupling matrix elements.
Using the above S-matrix, one is then able to calculate observables. The ensemble average $\langle S_{cc'}(\varepsilon)S^{\star}_{cc'}(\varepsilon')\rangle$ supplies the average conductance or cross section ($\varepsilon = \varepsilon'$), while the four-point function, $\langle S_{cc'}(\varepsilon_{1})S^{\star}_{cc'}(\varepsilon_{1})S_{cc'}(\varepsilon_{2})S^{\star}_{cc'}(\varepsilon_{2})\rangle$, furnishes the correlation function, characterized by the correlation width, of great importance in the study of chaotic quantum systems.\\

In application to conductance statistics in Quantum Dots with three or more terminals it is convenient to represent the $S$-matrix as,

\begin{equation}
S=\left(\begin{array}{ccc}r_{11} & t_{12} & t_{13} \\ t_{12}' & r_{22} & t_{23} \\ t_{13}' & t_{23}' & r_{33} \\
\end{array}\right),
\end{equation}
where $t_{ij}$ indicates the probability amplitude of transmission of the channel(s) contained in the terminal $ i $ for the channel(s) contained in the terminal  $ j$ and $ r_ {ii} $ denotes the probability amplitude of reflexions of the channels in the terminal $ i $.

The correlation function obtained from the above is invariably of a Lorentzian shape, in the case of a variation of the energy, or a square Lorentzian, in the case of a variation of the external parameter, $X$. Deviations from these limiting cases are expected and can be derived \cite{ref2,caio}. Calculation of the ensemble averages, is rather difficult, and only the case of the Gaussian Unitary Ensemble average has been performed analytically. Most applications of the above $S$-matrix, involves numerical simulations using random matrix generator. To obtain analytical results, an alternative method has been devised, based on the distribution of the $S$-matrix itself, which, being unitary, follows the Dyson circular ensemble. The Stub model is such an alternative \cite{brouwer,nos1,nos2}. It involves the use of the following parametrized form of the $S$-matrix.

\section{The Stub Model}

We assume that the particles dynamics is ballistic and ergodic, and we model the system statistical properties using the random matrix theory (RMT)-based stub model. Following references \cite{brouwer,nos2}, for particles with
spin, the scattering matrix S can be represented as a unitary matrix with quaternionic entries and is parameterized as
\begin{eqnarray}
{\cal S} (\varepsilon,{\cal B})={\bar {\cal S}} +{\cal P U}[1-{\cal K}^\dagger  {\cal R}(\varepsilon, {\cal X}) {\cal K} {\cal U}]^{-1}{\cal P}^\dagger. \label{SMatriz}
\end{eqnarray}
Here, $ {\cal U}$-matrix, $M\times M$, is the scattering matrix counterpart of an isolated quantum system, while ${\bar {\cal S}} $ is the average of the scattering matrix of the system ${\cal S}$, which has dimension $N\times N$.  The  $M$ stands for the number of resonances of the system, while $N=N_1+N_2+N_3$ is the total number of open channels. The universal regime requires $M \gg N$. The ${\cal K}$-matrix is a projection operator  of order $(M-N) \times M$, while $ {\cal P} $, of order $N \times M$, describes the channels-resonances couplings. Their explicit forms read $ {\cal K}_{i, j} = \delta_{i +N, j} $, $ {\cal P}_{i, j} = \textrm{diag} ( i \delta_{i, j} \sqrt{T_{1}}, i \delta_{i+N_1, j} \sqrt{T_{2}},i \delta_{i+N_1+N_2, j} \sqrt{T_{3}})$ and $ {\bar {\cal S}}_{i, j} = \textrm{diag} ( \delta_{i, j} \sqrt{1-T_{1}}, \delta_{i+N_1, j} \sqrt{1-T_{2}}, \delta_{i+N_1+N_2, j} \sqrt{1-T_{3}})$,  we are assuming the equivalent coupling for channels in the same terminal. The $ {\cal R}$-matrix (representing the external fields) is the stub model counterpart of order ($M-N \times M-N$) as described by \cite{nos2}. The parameter $X$ represents, e.g., the type of external applied magnetic field employed. The quantity $T = tr(t_{12}t_{12}^{\dagger})$.

\begin{figure*}[htp]
 \includegraphics[width=0.8\textwidth]{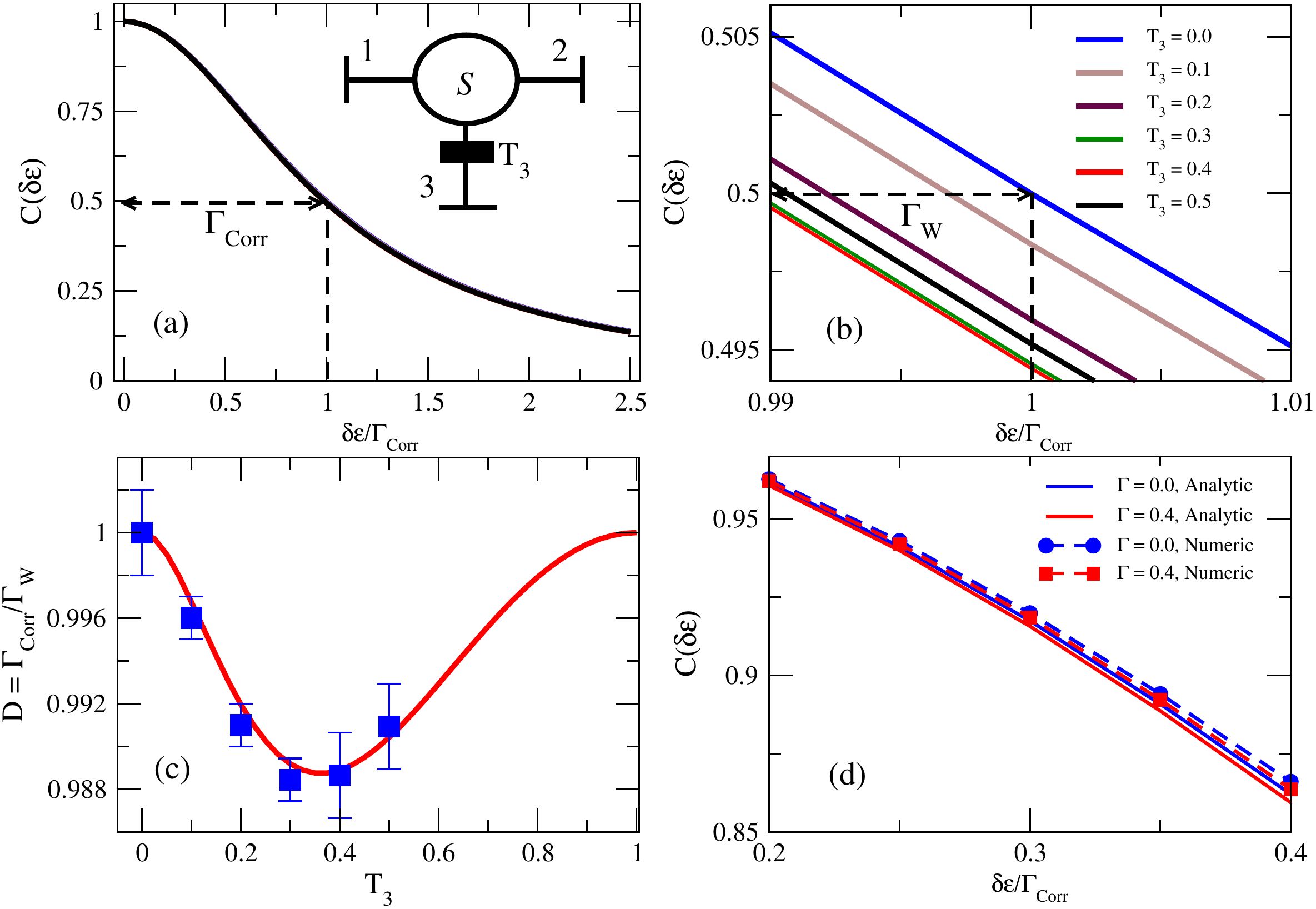}
\caption{\scriptsize Fig.(1a) show a typical Lorentzian behaviour of the correlation function and the correlation length $\Gamma_{corr}$ associated with a typical transport observable between the terminals $1$ and $2$ probed by another non-linear terminals $3$, as the inset indicates. Fig.(1b) shows a deviation from  the Weisskopf length in the semiclassical regime occasioned by the presence of another terminals. Fig.1(c) shows the ratio $D = \Gamma_{corr}/\Gamma_{W}$ as a function of $T_3$ exhibiting the attainment of the  Weisskopf length only for a completely open ($T_3 = 1$) or completely closed ($T_3 = 0$) extra-terminals lumped into an effective one. See text for details.. The continuous line is the analytical result and the dots are the results of the numerical Hamiltonian simulation using the $S$-matrix of Eq. (1). Fig.(1d) shows the deviation of the Weisskopf length analytically and numerically.}\label{grafico1}
 \end{figure*}

The calculation of the ensemble averages can be done in a closed form using the stub $S$-matrix above. Details of such calculations can be found in the original references. 
The ensemble average of the conductance, $Q = tr(t_{12}t'_{12})$, is easily calculated, \cite{Mello},  following the Ref. \cite{nos2} of a chaotic QS.  We obtain
\begin{eqnarray}
\langle {\cal Q}\rangle &=& \frac{{\cal T}_1 {\cal T}_2}{{\cal T}}\left[1-\frac{2-\beta}{\beta}\frac{{\cal T}_1 T_2+{\cal T}_2T_1}{{\cal T}^2}\right] ,\label{condg}
\end{eqnarray}
where ${\cal T}_i=N_i T_i$ is the total transmission coefficient for the terminal $i$ with equivalent channels and ${\cal T}={\cal T}_1+{\cal T}_2+{\cal T}_3$, and the $\varepsilon$-dependence disappear in the ensemble average. In the case of the compound nucleus, the quantity $Q$ becomes the cross section and the its ensemble average , similar to eq. (4), is known as the Hauser-Feshbach cross section \cite{HF}, with the factor $\left[1-\frac{2-\beta}{\beta}\frac{{\cal T}_1 T_2+{\cal T}_2T_1}{{\cal T}^2}\right] $ representing what is known as the elastic enhancement factor, which is known exactly at arbitrary overlap both for
orthogonal and unitary symmetry, \cite{verb,ref3}. Note that the number of channels, $N$,  in all the terminals is taken to be large, $N >> 1$, and thus we are in the semiclassical regime. Note further that the symmetry parameter $\beta$ takes on the values 1, in the Gaussian Orthogonal Ensemble, 2 in the Gaussian Unitary Ensemble, and 4 in the Gaussian Symplectic Ensemble.

\section{The Correlation Functions}
As for the correlation functions, the calculation using the stub model can be carried out as done in \cite{BHR2013}. In the following we merely write down the correlation functions for a chaotic Quantum Dot in the case of only two terminals, specified by $T_1$ and $T_2$ and and consider $T_1 = T_2 = T$, we find for the energy variation we find,

\begin{eqnarray}
\frac{C(\delta E)}{1/8 \beta}= \frac{3 T(2-T)-2}{1+(\delta E/\Gamma_{E})^2}+\frac{4 \left[ 1 + T (T-2) \right]}{\left[ 1+(\delta E/\Gamma_{E})^2 \right]^2}
\end{eqnarray}

where $\beta$ is a measure of the universality class to which $H$ pertains. It should be mentioned that magneto-conductance correlation functions, were calculated using the Hamiltonian approach, Eq. (1), and the study of a
relevant asymptotic expansion in inverse channel number was 
reported in \cite{weiden99}. 

For the variation of the external parameter, we consider three cases. An external perpendicular magnetic field, $B_{\perp}$, and external perpendicular magnetic filed acting on the spin-orbit interaction, the so-called Rashba-Drasselhaus field, $H_{\perp}$ \cite{Rashba,Dressel}, and a parallel magnetic field, $H_{\parallel}$, whose effect has been studied recently by \cite{Harvard},

\begin{eqnarray}
\frac{C(\delta X)}{1/8 \beta}= 
\frac {2 T(1-T)}{1+ (\delta X/\Gamma_{X})^2} + \frac{2+ T(3T-4) }{ \left[ 1+ ( \delta X/\Gamma_{X})^2 \right] ^{2}}
\end{eqnarray}

\begin{eqnarray}
C(\delta H_{\perp})=\frac {3\,T^{2}+4-4\,T }{ \left( {{ (\delta H_{\perp}/\Gamma_{H_{\perp}})^2}}+2
 \right) ^{2}}-\frac {T^{2}-2\,T }{{{(\delta H_{\perp}/\Gamma_{H_{\perp}}})^{2}+2}}
\end{eqnarray}
Finally for the case of a parallel magnetic field considered recently in \cite{Harvard}
\begin{eqnarray}
    \frac{C(\delta H_{\parallel})}{1/8 \beta}=\frac{T(2-T)-1}{1+(\delta H_{\parallel}/\Gamma_{H_{\parallel}})^2}
\end{eqnarray}
This last correlation function is new and deserves some discussion. It does not depend on the openness parameter, the transmission coefficient, $T$,  and it has a pure Lorentzian shape in contrast to the cares of perpendicular magnetic field. \\

It should be mentioned that a thorough study of the energy dependence and correlation length
in quantum chaotic scattering with many channels was performed in \cite{ref1}
A general relation was established there between fluctuations in
scattering and the distribution of complex energies (poles of the S
matrix). In particular, the correlation length was shown to be given
by the spectral gap in the pole distribution, and the deviations of
the gap from the (semiclassical) Weisskopf estimate [eq.(14) of the
present work] were analyzed in great detail there as well. 

\section{The Average Density of Maxima}
The important feature that characterizes these correlation functions is that they all (except the case of $H_{\parallel}$) deviate from pure Lorentzian or square Lorentzian shape, for an arbitrary value of the openness probability. The application to the case of the compound nucleus, Eq. ( 5), allows the investigation of the statistics of resonances both in the weak (isolated resonances) and strong (overlapping resonances) absorption cases, as well as in the intermediate cases. Several quantities can be obtained from the correlation functions. The average density of maxima in the fluctuating observable, is one of them. In Ref. \cite{RBHL2011}, this quantity was derived and analyzed for $C(E)$ and $C(X)$. For completeness we give below the main results  and extend them to other types of applied magnetic fields. 

\begin{eqnarray}
\left< \rho_{z} \right> &=&\frac{1}{2\pi}\sqrt{\frac{T_4}{T_2}} \label{dens} \\
T_2&=&-\frac{d^2}{d(\delta z)^2}C(\delta z)\bigg|_{\delta z=0}\nonumber\\
T_4&=&\frac{d^4}{d(\delta z)^4}C(\delta z)
\bigg|_{\delta z=0}.\nonumber
\end{eqnarray}

For the energy variation,

\begin{equation}
\left< \rho_{E} \right>=\frac{\sqrt{3}}{\pi \Gamma_{E}}\sqrt{\frac { 9\,{T }^{2}-18\,T+10}{ 5\,{T }^{2}-10\,T+6}} =\frac{\sqrt{3}}{\pi \Gamma_{corr}}
\end{equation}

For the case of an external perpendicular magnetic field,

\begin{equation}
\left< \rho_{X} \right>=\frac{\sqrt{3}}{\sqrt{2} \pi \Gamma_{X}}\sqrt{\frac{7\,{T }^{2}-10\,T+6 }{2\,{T }^{2}-3\,T+2}}
\end{equation}

For the Rashba-Dresselhaus field,

\begin{equation}
\left< \rho_{H_{\perp}} \right>=
\frac{\sqrt{3}}{2 \pi \Gamma_{h_{\perp}}} \sqrt{\frac {7\,{T }^{2}-10\,T+6}{2\,{T }^{2}-3\,T+2}}
\end{equation}

Interestingly for the case of a parallel magnetic field, the result is independent of the openness parameter and is identical to the Brink-Stephen \cite{Brink} one,
\begin{equation}
\left< \rho_{H_{\parallel}} \right> = \frac{\sqrt{3}}{\pi \Gamma_{H_{\parallel}}}  \approx \frac{0.55}{\Gamma_{H_{\parallel}}}
\end{equation}

\section{The Correlation Length and Weisskopf's Estimate}
The important point to emphasize here is that both the correlation function and the average density of maxima are characterized, for a fixed value of the openness probability, by a single quantity, the correlation width. In the energy variation case, this width is the inverse of the dwell time and is usually estimated using the Weisskopf expression \cite{Weisskopf},

\begin{equation}
\Gamma_{E} \approx \Gamma_{W} = \frac{\overline{\Delta}}{2\pi}\sum_{c} T_{c}
\end{equation}
where, $\overline{\Delta}$ is the average spacing between the resonances in the chaotic system, and the sum extends over all the open channels reached through the transmission coefficient, $T_{c}$.
Deviation from the Weisskopf estimate was calculated in \cite{Richter} using the $S$-matrix of Eq. ( 1). With the help of a random matrix generator, these authors calculated numerically the transmittance correlation function for microwave resonators, and obtained the correlation width, $\Gamma_{corr}$ as the width at half maximum. They found for the ratio $D = \Gamma_{corr}/\Gamma_{W}$ vs. $\sum_{c} T_{c}$, values that reach up to 1.1. In our work here, we can read out the change in the correlation width as,

\begin{equation}
D = \frac{\Gamma_{corr}}{\Gamma_{W}} = \sqrt{\frac{ 5\,{T}^{2}-10\,T+6}{ 9\,{T }^{2}-18\,T+10}}
\end{equation}
For almost closed system, $T << 1$, the deviation reaches the value $D = \sqrt{3/5}$ = 0.81. Of course in the other limit of a completely open Quantum Dot, $T \approx$ 1, the ratio $D$ attains the value of unity as expected.
.

It is interesting to generalize our result to the case of two terminals in the presence of other terminals lumped together as an effective terminal 
characterized by $T_{3}$, and study the variation of the correlation width as a function of 
$T_3$. Thus, we extend the stub method for the new calculation of the general correlation function, including the quantum interference terms. 
Using the same stub ${\cal S}$-matrix described above, together with the typical large number of diagrams for the correlation function 
\cite{brouwer}, we calculate the averages of products of two observables, $Q(E) Q(E')$, and obtain after setting ${\cal T}_1 = {\cal T}_2 = N$, 
and $N_1 = N_2 = N_3 = N$. Note that in this case of more than two terminals, the result we have obtained are valid for $T_1 = T_2 = 1$, 
to get.

\begin{eqnarray}
\frac{{\cal C}(\delta E)}{1/\beta}&=&\frac{\left[{\cal A}_1(T_3)-8N^2{\cal T}_3^2(1-T_3)^2\right]}{N^4(2 +T_3)^6\left[1+(\delta E/\Gamma_W)^2\right]}\nonumber\\
&+&\delta_{2\beta}\frac{{\cal T}_3 {\cal A}_{2}(T_3)}{N^3(2 +T_3)^6 \left[1+(\delta E/\Gamma_W)^2\right]}\nonumber\\
&+&\frac{8 {\cal T}_3^2(1-2T_3+T_3^2)}{(2 + T_3)^6\left[1+(\delta E/\Gamma_W)^2\right]^2}
 \label{geral}
\end{eqnarray}
 and,

\begin{eqnarray}
\frac{{\cal A}_{1}(T_3)}{N^4} &\equiv & A_{1}(T_3) =T_3^4+2 (2p +T_3)(2+T_3)T_3^{3}\nonumber\\
&+&\left[2T_3^{2}-4(2 + T_3)^{2}T_3\right.\nonumber\\
&+&\left.16(2 + T_3)^2 \right ]T_3^2\nonumber\\
&+&2(2p + T_3)\left[-4T_3^{2}(2 + T_3)\right. \nonumber\\
&+&\left.12 (2 + T_3)^2\right]T_3+8(2 +T_3)^4\nonumber\\
\frac{{\cal A}_{2}(T_3)}{N^3}&\equiv &A_{2}(T_3) = T_{3}(2 +T_3)^2\left[2(2 +T_3)+ 1\right]\nonumber\\
\end{eqnarray}

The Eq.(\ref{geral}) shows convincingly that the correlation function is not a Lorentzian on the Weisskopf width scale, which is violated and no longer the transport correlation width. The Fig. (1b) shows the modification of the Weisskopf length as a function of the $T_3$. On the another hand, the amplitude of the universal fluctuations (variance) ${\textrm var}[{\cal Q}]= {\cal C}(0)=1/\beta$ is maintained regardless of the additional terminals. The Eq.(\ref{geral}) can be written in Lorentzian form, using some algebra, as in the following
\begin{equation}
    \frac{C(\delta E)}{1/\beta}=\frac{1}{1+\left( \delta E/\Gamma_{corr}\right)^2}, \;\; \Gamma_{corr}\equiv \Gamma_W D\label{lorentziana}
\end{equation}
with
\begin{eqnarray}
D&\equiv& \frac{\sqrt{64T_3^{4}(1-T_3)^4+\left[A_{1}(T_3)+\delta_{2\beta}T_{3}A_{2}(T_3)\right]^2}}{A_{1}(T_3)+\delta_{2\beta}T_{3} A_{2}(T_3)}\nonumber\\&-&\frac{8T_3^{2}(1-T_3)^2}{A_{1}(T_3)+\delta_{2\beta}T_{3}A_{2}(T_3)} \label{lorentzianaC}
\end{eqnarray}
which is highly non-liner  as a function of $T_3$. The form of the Eqs.(\ref{lorentziana}) and (\ref{lorentzianaC}) show analytically the effect of the presence of more terminals on the correlation properties of the other two terminals. It simulates absorption of the flux in the two-terminal subsystem. The non-linear effect disappears in  the limits $ T_3 = 1$ (ideal) and $ T_3 \rightarrow 0 $ (opaque) for which $D \rightarrow 1$, and the correlation length approaches the  Weisskopf length, as Fig.(1c) shows clearly. These results were verified by a numerical simulations with random matrix generator using the $S$-matrix of Eq. (1). 
\begin{figure}[h!]
	\centering
	        \vskip-0.4cm
		\includegraphics[width=\columnwidth]{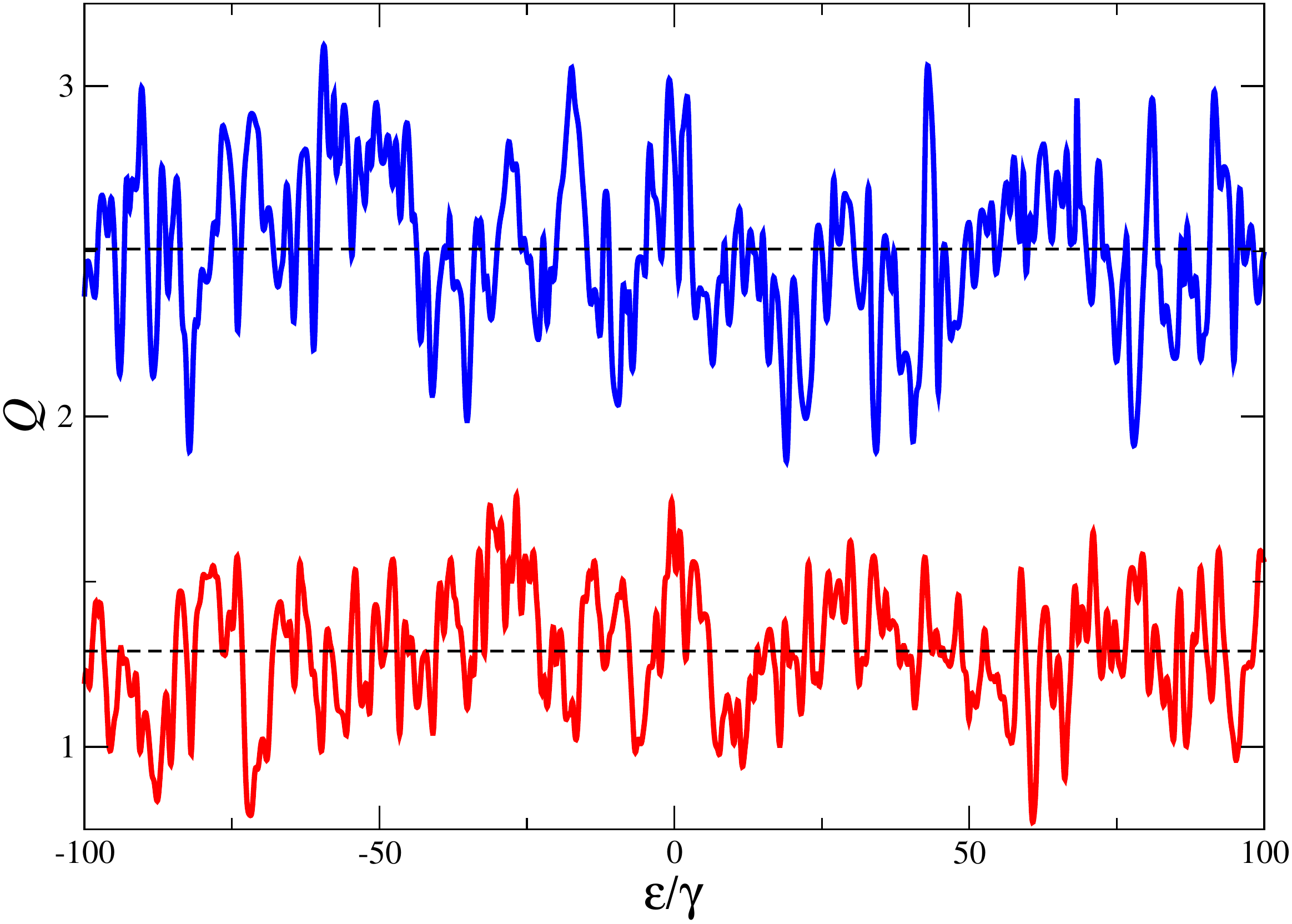}
	\caption{Typical transport observable ${\cal Q}$ as a function of $\varepsilon/\Gamma_W$ for $N=50$ open channels coupled to $5 \cdot 10^3$ resonances. For a single realization of $H$ in both fluctuations and the same observable ${\cal Q}$, the top indicates a single measure in the absence of the extras-terminals and the bottom indicates extra-terminals with non-linear coupling $T_3=0.3$.}
	\label{fig:TvsEN=5}
\end{figure}

We perform a numerical simulation using the Hamiltonian model of Eq. (1) with a configuration of $50$ open channels and $5 \cdot 10^3$ resonances in an energy range $ \Delta \varepsilon / \gamma \in [-100,100]$. As shown in Fig.[2], the transport observable ${\cal Q}$ is ``experimentally" obtained for the same scattering QS. For such single QS the ${\cal Q}$ is largely affected in the presence of the extra non-linear terminals in two forms. Firstly, the reference line of fluctuations is suppressed (from the top to the bottom plot of Fig.[2]), as can be expected using Eq.(4). Secondly, the number of maxima increases from 99 to 101 in the same energy interval, confirming nicely our analytical findings. A single experimental trace can confirm the violation of Weisskopf width in the presence of the extra terminals.

Thus, we find excellent agreement between our analytical calculation and the numerical one. In a way we are also confirming the results of \cite{Richter} obtained for transmittance in a microwave cavity.

\section{Conclusions} In this paper, we analytically calculate the correlation functions for chaotic systems using the Quantum Chaotic Scattering theory. We show that in the case of energy variation and in the variation of an external magnetic field, these functions deviate significantly from the expected Lorentzian and square Lorentzian shapes. The parameter that measures this deviation is identified as the transmission coefficient which acts as the "openness" probability. We further identify the deviation of the correlation width from the Weisskopf width for an arbitrary observable of the quantum transport. The stub calculation of the ensemble average of the products of four scattering matrices allows finding this deviation, of the order of $10\%$, and reveals the topological effects of other terminals on the chaotic quantum transport. The consequences of our results include the increase of the dwell time of an arbitrary chaotic scattering system. The results are general and applicable to any scattering amplitudes between two terminals within the diagrammatic approach of the stub model, which is an 1/N expansion method. Therefore the deviation of the correlation width from the Weisskopf estimate can occur in spin and/or charge channels for electronic nanostructures, the transmittance of antennas, sub-lattices and/or sub-valleys channels for graphene flakes \cite{graphene}, etc. Technological applications of the results includes artificial atomic clocks by means of dwell time modifications and cryptography on transport observables.

This work is supported in part by the Brazilian funding agencies CAPES, CNPq, FACEPE, FAPESP, the Instituto Nacional de Ci\^{e}ncia e Tecnologia de Informa\c{c}\~{a}o Qu\^{a}ntica-MCT, and the CEPID-CEPOF Project.

\end{document}